\documentstyle[12pt]{article}
\catcode`\@=11
\@addtoreset{equation}{section}
\def\theequation{\arabic{section}.\arabic{equation}}
\def\section{\@startsection{section}{1}{\z@}{3.5ex plus 1ex minus
   .2ex}{2.3ex plus .2ex}{\large\bf}}
\def\thesection{\arabic{section}}

\def\appendix{\setcounter{section}{0}
        \def\thesection{Appendix}
        \def\theequation{\Alph{section}.\arabic{equation}}}

\def\eqnarray{\let\@currentlabel=\theequation\refstepcounter{equation}
    \global\@eqnswtrue
    \global\@eqcnt\z@\tabskip\@centering\let\\=\@eqncr
    $$\halign to \displaywidth\bgroup\@eqnsel\hskip\@centering
      $\displaystyle\tabskip\z@{##}$&\global\@eqcnt\@ne
       \hfil${{}##{}}$\hfil
      &\global\@eqcnt\tw@ $\displaystyle\tabskip\z@{##}$\hfil
       \tabskip\@centering&\llap{##}\tabskip\z@\cr}
\def\lefteqn#1{\hbox to 4\arraycolsep{$\displaystyle #1$\hss}}
\begin{document}
%%%%%%%%%%%%%%%%%%%%%%%%%%%%%%%%%%%%%%%%%%%%%%%%%%%%%%%%%%%%%%%%%%%%%%%%%%%
%     C I T E . S T Y
%     compressed lists of numerical citations: [11-16]
%     see also OVERCITE.STY and DRFTCITE.STY
%
%     Copyright (C) 1989-1992 by Donald Arseneau
%     These macros may be freely transmitted, reproduced, or modified for
%     non-commercial purposes provided that this notice is left intact.
%
%
%  \@citen contains the code that parses the list of names, ignoring
%  spaces after commas, writes the aux file \citation, and formats the
%  number list.  \citen can be used by itself to give citation numbers
%  without the other formatting; e.g., "See also ref.~\citen{junk}."
%
\def\citen#1{%
\edef\@tempa{\@ignspaftercomma,#1, \@end, }% ignore spaces in parameter list
\edef\@tempa{\expandafter\@ignendcommas\@tempa\@end}%
\if@filesw \immediate \write \@auxout {\string \citation {\@tempa}}\fi
\@tempcntb\m@ne \let\@h@ld\relax \let\@citea\@empty
\@for \@citeb:=\@tempa\do {\@cmpresscites}%
\@h@ld}
%
% for ignoring spaces in the input:
\def\@ignspaftercomma#1, {\ifx\@end#1\@empty\else
   #1,\expandafter\@ignspaftercomma\fi}
\def\@ignendcommas,#1,\@end{#1}
%
% For each citation, check if it is defined, if it is a number, and
% if it is a consecutive number that can be represented like 3-7.
%
\def\@cmpresscites{%
 \expandafter\let \expandafter\@B@citeB \csname b@\@citeb \endcsname
 \ifx\@B@citeB\relax % undefined
    \@h@ld\@citea\@tempcntb\m@ne{\bf ?}%
    \@warning {Citation `\@citeb ' on page \thepage \space undefined}%
 \else%  defined
    \@tempcnta\@tempcntb \advance\@tempcnta\@ne
    \setbox\z@\hbox\bgroup % check if citation is a number:
    \ifnum\z@<0\@B@citeB \relax
       \egroup \@tempcntb\@B@citeB \relax
       \else \egroup \@tempcntb\m@ne \fi
    \ifnum\@tempcnta=\@tempcntb % Number follows previous--hold on to it
       \ifx\@h@ld\relax % first pair of successives
          \edef \@h@ld{\@citea\@B@citeB}%
       \else % compressible list of successives
%         % use \hbox to avoid easy \exhyphenpenalty breaks
          \edef\@h@ld{\hbox{--}\penalty\@highpenalty \@B@citeB}%
       \fi
    \else   %  non-successor--dump what's held and do this one
       \@h@ld \@citea \@B@citeB \let\@h@ld\relax
 \fi\fi%
 \let\@citea\@citepunct
}
%
%%    To put space after the comma, use:
\def\@citepunct{,\penalty\@highpenalty\hskip.13em plus.1em minus.1em}%
%%    For no space after comma, use:
%% \def\@citepunct{,\penalty\@highpenalty}%
%%
%
%  Make \@citex refer to \citen:
%
\def\@citex[#1]#2{\@cite{\citen{#2}}{#1}}%
%
%  Replacement for \@cite.  Give one normal space before the citation,
%  set high penalties for linebreaks,
%
\def\@cite#1#2{\leavevmode\unskip
  \ifnum\lastpenalty=\z@ \penalty\@highpenalty \fi % highpenalty before
  \ [{\multiply\@highpenalty 3 #1% % triple-highpenalties within list
      \if@tempswa,\penalty\@highpenalty\ #2\fi % and before note.
    }]\spacefactor\@m}
\let\nocitecount\relax  % in case \nocitecount was used for drftcite
%
%%%%%%%%%%%%%%%%%%%%%%%%%%%%%%%%%%%%%%%%%%%%%%%%%%%%%%%%%%%%%%%%%%%%%%%%%%
\begin{titlepage}
\begin{flushright}
UCD-98-11\\
DFTT-33-98\\
gr-qc/9807087\\
July 1998\\
\end{flushright}
\vspace{.5in}
\begin{center}
{\Large\bf
 The Quantum Modular Group\\[1ex]
 in (2+1)-Dimensional Gravity}\\
\vspace{.4in}
{S.~C{\sc arlip}\footnote{\it email: carlip@dirac.ucdavis.edu}\\
       {\small\it Department of Physics}\\
       {\small\it University of California}\\
       {\small\it Davis, CA 95616}\\{\small\it USA}}\\
\vspace{1ex}
{\small and}\\
\vspace{1ex}
{J.\ E.~N{\sc elson}\footnote{\it email: nelson@to.infn.it}\\
       {\small\it Dipartimento di Fisica Teorica }\\
       {\small\it Universit\`a degli Studi di Torino}\\
       {\small\it via Pietro Giuria 1, 10125 Torino}\\{\small\it Italy}}

\end{center}

\vspace{.5in}
\begin{center}
\begin{minipage}{4.6in}
\begin{center}
{\large\bf Abstract}
\end{center}
{\small
The role of the modular group in the holonomy representation of
(2+1)-dimensional quantum gravity is studied.  This representation
can be viewed as a ``Heisenberg picture,'' and for simple topologies,
the transformation to the ADM ``Schr{\"o}dinger picture'' may be
found.  For spacetimes with the spatial topology of a torus, this
transformation and an explicit operator representation of the mapping
class group are constructed.  It is shown that the quantum modular
group splits the holonomy representation Hilbert space into physically
equivalent orthogonal ``fundamental regions'' that are interchanged
by modular transformations.
}
\end{minipage}
\end{center}
\end{titlepage}
\addtocounter{footnote}{-2}

\section{Introduction \label{sec1}}

Over the past few years, it has become apparent that (2+1)-dimensional
general relativity can provide a valuable setting in which to explore
some of the fundamental issues of realistic (3+1)-dimensional quantum
gravity \cite{Carlip_bk}.  As a diffeomorphism-invariant theory of spacetime
geometry, the (2+1)-dimensional model shares the conceptual framework
of ordinary (3+1)-dimensional gravity.  At the same time, however, the
reduction in the number of dimensions greatly simplifies the structure:
(2+1)-dimensional general relativity has only a finite number of physical
degrees of freedom, and quantum field theory is effectively reduced to
quantum mechanics.

At least fifteen different approaches to quantizing (2+1)-dimensional
general relativity have been developed over the past decade.  Two that
have received special attention are reduced phase space quantization,
starting with the ADM formalism and the York time-slicing \cite{HosNak,%
Mon,Fujiwara}, and a set of techniques that take Chern-Simons holonomies
as the fundamental observables \cite{Achu,Witten,observables,dirac,%
ordering,NR2,NR1,NR3,NR5,NR4,NR0,NRZ,Unruh}.  Both approaches to
quantization are well understood for the simplest topologies, and
in particular for spacetimes with the spatial topology of a torus,
$M\approx{\bf R}\times T^2$.  For these topologies, the two techniques
yield complementary information about the quantum behavior, and a
comparison has offered valuable insights into both \cite{dirac,%
CarlipNelson,CarlipNelson2}.

One persistent problem has, however, plagued this program.  In addition
to the usual ``small'' diffeomorphisms, the torus $T^2$ admits ``large''
diffeomorphisms, diffeomorphisms that cannot be continuously deformed
to the identity.  In ADM quantization, the natural configuration space
is Teichm{\"u}ller space, and the group of large diffeomorphisms---the
modular group---has a well-understood and well-behaved action on this
space.  As a consequence, standard mathematical results allow us to
construct invariant (or more general ``covariant'') wave functions
\cite{ordering,Puzio}.  In the holonomy representation, on the other
hand, the modular group does {\em not\/} act nicely (i.e., properly
discontinuously) on the natural configuration space, and the construction
of invariant wave functions is much more problematic \cite{Louko,%
Giulini,Peldan}.  Since the two approaches are supposed to be equivalent,
this mismatch is a cause for concern.

In this paper, we resolve this problem by explicitly constructing a
transformation between the two representations.  In the ADM
representation, the modular group splits the configuration space into
fundamental regions that are interchanged by the action of the group, and
an invariant wave function can be defined by giving its value on a single
fundamental region.  In the holonomy representation, no invariant wave
functions exist.  But we shall see that the {\em Hilbert space\/} now
splits into orthogonal ``fundamental regions'' that are interchanged by
a unitary action of the modular group.  Each of these subspaces is
equivalent, and each is equivalent to the ADM Hilbert space of invariant
(technically, weight-$1/2$) wave functions.  The choice of one such
subspace is the discrete analog of a choice of gauge, and once such a
choice is made, the conflict between the two quantizations disappears.

\section{Two Quantizations \label{sec2}}

We start with a very brief review of the two approaches to quantization
described in the introduction, focusing on the torus universe ${\bf R}\times
T^2$.  For simplicity, we shall consider only a negative cosmological
constant, $\Lambda = - 1/\alpha^2$.  Details can be found in references
\cite{CarlipNelson,CarlipNelson2} and \cite{Carlip_bk}.

To construct an ADM quantization, we first foliate the spacetime
${\bf R}\times T^2$ by time slices of constant mean (extrinsic) curvature
$k$ \cite{York}.  The fixed value of $k$ on a slice then serves as a
time coordinate.  The geometry of each $T^2$ slice is determined up to
a conformal factor by a complex modulus $\tau = \tau_1 +i\tau_2$,
\begin{equation}
d\sigma^2 = e^{2\lambda}{\tau_2}^{-1}\left| dx + \tau dy \right|^2 .
\label{bb5}
\end{equation}
It may be shown that the conformal factor is fixed by the Hamiltonian
constraint, leaving a physical phase space parametrized by the variables
$\tau_1$ and $\tau_2$ and their conjugate momenta $p^1$ and $p^2$, or
equivalently by complex variables $\tau$ and $p=p^1 +i p^2$.  Evolution
in constant mean curvature time $k$ is generated by an effective
Hamiltonian that is just the spatial volume \cite{HosNak,Mon},
\begin{equation}
H = \int_{T^2}d^2x \sqrt{^{(2)}g}
  = {1\over\sqrt{k^2 - 4\Lambda}}{\tilde H}, \qquad
{\tilde H} = \tau_2\sqrt{p \bar p}.
\label{11}
\end{equation}
The quantity $\tilde H$ may be recognized as the square of the momentum
$p$ with respect to the Poincar\'e (constant negative curvature) metric
\begin{equation}
d\ell^2 = \tau_2{}^{-2}d\tau d\bar\tau  ,
\label{bb13a}
\end{equation}
the standard metric on the torus moduli space.  The basic Poisson brackets
are
\begin{equation}
\left\{ \tau, \bar p\right\} = \left\{ \bar \tau, p\right\} = 2 ,\quad
\left\{ \tau, p\right\} = \left\{ \bar \tau, \bar p\right\} = 0,
\label{bb14}
\end{equation}
and the reduced Einstein action becomes
\begin{equation}
I_{\hbox{\scriptsize \it Ein}}
  = \int dk \left( p^\alpha {d{\tau_\alpha}\over dk} - H(\tau,p,k)
\right).
\label{bb10}
\end{equation}

The reduction to the variables $\tau$ and $p$ eliminates the ``small''
diffeomorphisms, but a group of ``large'' diffeomorphisms, the modular
group, remains.  One set of generators of this group consists of
two transformations $S$ and $T$, which act classically as
\begin{eqnarray}
&S&: \tau\rightarrow -\tau^{-1} ,\quad
     p\rightarrow \bar \tau^2 p,\nonumber\\
&T&: \tau\rightarrow \tau+1 ,\quad p\rightarrow p
\label{bb15}
\end{eqnarray}
and satisfy the identities
\begin{equation}
S^2=1 ,\quad (ST)^3 =1 .
\label{modid}
\end{equation}
These transformations leave the Hamiltonian (\ref{11}) and Poisson brackets
(\ref{bb14}) invariant.

The reduced phase space action (\ref{bb10}) is equivalent to that of a
finite-dimensional mechanical system with a complicated Hamiltonian.
We know, at least in principle, how to quantize such a system: we simply
replace the Poisson brackets (\ref{bb14}) with commutators,
\begin{equation}
\left[ \hat \tau_\alpha, \hat p^\beta \right] = i\hbar\delta^\beta_\alpha ,
\label{da1}
\end{equation}
represent the momenta as derivatives,
\begin{equation}
p^\alpha = {\hbar\over i}{\partial\ \over\partial \tau_\alpha} ,
\label{da2}
\end{equation}
and impose the Schr\"odinger equation
\begin{equation}
i\hbar{\partial\psi(\tau,k)\over\partial k} = \hat H\psi(\tau,k) ,
\label{da3}
\end{equation}
where the Hamiltonian $\hat H$ is obtained from (\ref{11}) by some
suitable operator ordering.

One fundamental problem is hidden in this last step: it is not at all
obvious how one should define $\hat H$ as a self-adjoint operator on an
appropriate Hilbert space.  In particular, ${\hat\tau}_2$ and $\hat p$
do not commute, so the operator ordering in $\hat H$ is not unique.  The
simplest choice is that of equation (\ref{11}), for which the Hamiltonian
becomes
\begin{equation}
\hat H = {\hbar\over\sqrt{k^2-4\Lambda}}\,\Delta_0^{1/2} ,
\label{da4}
\end{equation}
where $\Delta_0$ is the ordinary scalar Laplacian for the constant
negative curvature moduli space characterized by the metric (\ref{bb13a}).
Other orderings exist, but they are severely restricted by the requirement
of diffeomorphism invariance: eigenfunctions of $\hat H$ should transform
under a unitary representation of the modular group.\footnote{This
representation is usually assumed to be one dimensional, but it may
be permissible to consider higher-dimensional representations and
multicomponent wave functions.}  The representation theory of the modular
group has been studied extensively \cite{Fay,Maass,Rankin,Terras}; if we
restrict our attention to one-dimensional representations, the possible
Hamiltonians are all of the form (\ref{da4}), but with $\Delta_0$
replaced by\footnote{See \cite{ordering} for details of the required
operator orderings.}
\begin{equation}
\Delta_n = -\tau_2^{\ 2}\left( {\partial^2\ \over\partial\tau_1{}^2} +
  {\partial^2\ \over\partial\tau_2{}^2}\right)
  + 2in\, {\tau_2}{\partial\ \over\partial \tau_1} + n(n+1) , \quad
2n\in{\bf Z} .
\label{da5}
\end{equation}
The operator $\Delta_n$ is the weight $n$ Maass Laplacian, and the
corresponding eigenfunctions, Maass forms of weight $n$, have been
discussed in considerable detail in the mathematical literature
\cite{Fay,Maass,Rankin,Terras}.  Note that when written in terms of the
momentum $p$ of equation (\ref{da2}), the $\Delta_n$ differ from each other
by terms of order $\hbar$, as expected for operator ordering ambiguities.
Nevertheless, the choice of ordering can have drastic effects on the
physics: the spectra of the various Maass Laplacians are very different.

This ambiguity can be viewed as a consequence of the structure of the
classical phase space.  The torus moduli space is not a manifold, but
rather has orbifold singularities, and quantization on an orbifold
is generally not unique.  Since the space of solutions of the Einstein
equations in 3+1 dimensions has a similar orbifold structure \cite{Fischer},
we might expect a similar ambiguity in realistic (3+1)-dimensional
quantum gravity.

A potentially more serious ambiguity in this approach to quantization
comes from the classical treatment of the time slicing.  The choice of $k$
as a time variable is rather arbitrary, and it is not at all clear that
a different choice would lead to the same quantum theory.  The danger
of making a ``wrong'' choice is illustrated by the classical solution
(\ref{c1})--(\ref{c2}) described below: another standard slicing uses
$\sqrt{{}^{(2)}g}$ as time, but it is evident that when $\Lambda<0$,
$\sqrt{{}^{(2)}g}$ is not even a single-valued function of $k$.

A possible resolution of this problem is to treat the holonomy
representation as fundamental.  In this first-order ``frozen time''
approach, the basic observables give a time-independent description of the
entire spacetime geometry.  There is no Hamiltonian, no time development,
and hence no need to choose a time slicing.  If we can establish a
relationship between the $(\hat\tau,\hat p)$ and suitable operators in
the first-order formalism, we can convert the problem of time slicing
into one of defining the appropriate physical operators.  Different
choices of slicing would then merely require different operators to
represent moduli, and not different quantum theories.

The holonomy representation \cite{NR2,NRZ} starts with the Chern-Simons
formulation of (2+1)-dimensional gravity \cite{Achu,Witten}, and chooses
as fundamental variables the traces of the Chern-Simons holonomies
around a set of noncontractible curves $\{\gamma_a\}$.  For $\Lambda<0$,
the relevant gauge group is a product group $\hbox{SL}(2,{\bf R})\otimes%
\hbox{SL}(2,{\bf R})$ coming from the decomposition of the spinor group of
$\hbox{SO}(2,2)$ (the anti-de Sitter group), and one obtains two real,
independent sets of traces $R_a^{\pm}$ \cite{NR1,NRZ}.

For the torus, the algebra is simplest if we consider holonomies around
three curves: two circumferences $\gamma_1$ and $\gamma_2$ and a third
curve $\gamma_{12}= \gamma_1\cdot\gamma_2$, where the dot represents
composition of curves or multiplication of homotopy classes.  The
holonomies then satisfy the nonlinear Poisson bracket algebra
\begin{equation}
\{R_1^{\pm},R_2^{\pm}\}=\mp{1\over {4\alpha}}(R_{12}^{\pm}-
 R_1^{\pm}R_2^{\pm}) \quad \hbox{\it and cyclical permutations}.
\label{b7}
\end{equation}
The six holonomies $R^\pm_{1,2,12}$ provide an overcomplete description
of the spacetime geometry of ${\bf R}\!\times\!T^2$, which is completely
characterized by two complex parameters $\tau$ and $p$.  To remove
this overcompleteness, consider the cubic polynomials
\begin{equation}
F^{\pm}=1-(R_1^{\pm})^2-(R_2^{\pm})^2-(R_{12}^{\pm})^2 +
 2 R_1^{\pm}R_2^{\pm}R_{12}^{\pm} .
\label{b9}
\end{equation}
These polynomials have vanishing Poisson brackets with all of the
traces $R_a^{\pm}$, are cyclically symmetric in the $R_a^{\pm}$, and
vanish classically by the $\hbox{SL}(2,{\bf R})$ Mandelstam identities;
setting $F^\pm = 0$ removes the redundancy.

The Poisson algebra (\ref{b7}) and its generalization \cite{NR3} to more
complicated spatial topologies can be quantized for any value of the
cosmological constant.  For a generic topology, one obtains an abstract
quantum algebra \cite{NR1,NR0}.  For genus $1$ with $\Lambda < 0$, the
quantum theory has been worked out quite explicitly.

There are, in fact, two closely related theories: one can either quantize
the algebra and then determine a represention, or first choose a classical
representation and then quantize.  For the first choice, one replaces
the classical Poisson brackets $\{\,,\,\}$ by commutators $[\,,\,]$,
\begin{equation}
\{x,y\} \to {1\over i\hbar}[x,y] ,
\end{equation}
and replaces products in (\ref{b7}) by symmetrized products,
\begin{equation}
xy \to {1 \over 2} (xy +yx) .
\end{equation}
The resulting operator algebra is given by
\begin{equation}
\hat R_1^{\pm}\hat R_2^{\pm}e^{\pm i \theta}
  - \hat R_2^{\pm}\hat R_1^{\pm} e^{\mp i \theta}=
  \pm 2i\sin\theta\, \hat R_{12}^{\pm} \quad \hbox{\it and cyclical
  permutations}
\label{za}
\end{equation}
with
\begin{equation}
\tan\theta= -{\hbar/ 8\alpha} .
\label{za1}
\end{equation}
The algebra (\ref{za}) is not a Lie algebra, but it is related to the
Lie algebra of the quantum group $\hbox{SU}(2)_q$ \cite{NR5,NRZ}, where
$q=\exp{4i\theta}$, and where the cyclically invariant $q$-Casimir is
the quantum analog of the cubic polynomial (\ref{b9}),
\begin{equation}
\hat F^{\pm}(\theta)
  = {\cos}^2\theta- e^{\pm 2i\theta} \left( (\hat R_1^\pm)^2+
(\hat R_{12}^\pm)^2\right) -e^{\mp 2i\theta} (\hat R_2^\pm)^2
 + 2e^{\pm i\theta}\cos\theta \hat R_1^\pm \hat R_2^\pm \hat R_{12}^\pm .
\end{equation}
It may be checked that traces $\hat R_a$ satisfying (\ref{za}) can
be represented by \cite{NR1,CarlipNelson,CarlipNelson2}
\begin{equation}
\hat R_1^\pm = \sec\theta \,\cosh{{\hat r}_1^\pm\over2} , \quad
\hat R_2^\pm = \sec\theta \,\cosh{{\hat r}_2^\pm\over2} , \quad
\hat R_{12}^\pm = \sec\theta \,\cosh{({\hat r}_1^\pm+{\hat r}_2^\pm)\over2} ,
\label{c6}
\end{equation}
where the operators ${\hat r}_1^\pm$, ${\hat r}_2^\pm$  have the
commutators
\begin{equation}
[{\hat r}_1^{\pm}, {\hat r}_2^{\pm}] =  \pm {8i \theta}
\qquad [{\hat r}^+_a,{\hat r}^-_b] = 0 .
\label{commr}
\end{equation}

Alternatively, we could start with a classical representation of the
holonomies $R_a^\pm$ analogous to the $\hbar\rightarrow 0$ limit of
(\ref{c6}),
\begin{equation}
R_1^\pm = \cosh{r_1^\pm\over2} , \quad
R_2^\pm = \cosh{r_2^\pm\over2} , \quad
R_{12}^\pm = \cosh{(r_1^\pm+r_2^\pm)\over2} ,
\label{cc6}
\end{equation}
which will satisfy the algebra (\ref{b7}) provided the parameters
$r^{\pm}_a$ satisfy
\begin{equation}
\{r_1^\pm,r_2^\pm\}=\mp {1/\alpha}, \qquad \{r^+_a,r^-_b\}=0 .
\label{a1}
\end{equation}
In this case the cubic polynomials (\ref{b9}) are identically zero.
Quantization of (\ref{a1}) then gives
\begin{equation}
[\hat r_1^\pm, \hat r_2^\pm] = \mp {i\hbar/\alpha} .
\label{dc1}
\end{equation}
{}From (\ref{za1}), we see that this expression differs from (\ref{commr})
by terms of order $\hbar^3$.  For the rest of this paper we will consider
only the commutators (\ref{dc1}); the alternative quantization (\ref{commr})
can be obtained by a fairly simple rescaling.

In either approach, the modular group acts both classically and quantum
mechanically on the holonomy parameters as
\begin{eqnarray}
&S&: \hat r_1^{\pm}\rightarrow \hat r_2^{\pm},\quad
    \hat r_2^{\pm}\rightarrow -\hat r_1^{\pm},\nonumber\\
&T&: \hat r_1^{\pm}\rightarrow \hat r_1^{\pm} + \hat r_2^{\pm},\quad
    \hat r_2^{\pm}\rightarrow \hat r_2^{\pm},
\label{28}
\end{eqnarray}
and satisfies
\begin{equation}
S^2=-1 ,\quad (ST)^3 =1 ,
\label{modid1}
\end{equation}
as is appropriate for a spinor representation. The action leaves invariant
the Poisson brackets (\ref{a1}) and the commutators (\ref{dc1}).

It will later prove useful to have an explicit representation of the
${\hat r}^\pm_a$ as multiplicative and differential operators, analogous
to the representation (\ref{da2}) in ADM quantization.  An obvious choice
is to take the ${\hat r}_2^\pm$ as our configuration space variables, and
the ${\hat r}_1^\pm$ as momenta.  To simplify future algebra, though, it
is useful to pick instead a pair of linear combinations of the
${\hat r}_2^\pm$ to parametrize our configuration space.  Let $t$
denote the time coordinate in proper time gauge, related to the York
time $k$ by equation (\ref{c5}) below, and define
\begin{eqnarray}
 u &=& (\sin{2t \over {\alpha}})^{-{1\over 2}}\,(r_2^-e^{it/\alpha} +
r_2^+e^{-{it/\alpha}}), \nonumber \\
{\bar u} &=& (\sin{2t \over {\alpha}})^{-{1\over 2}}\,
(r_2^-e^{-{it/\alpha}} + r_2^+e^{it/\alpha}) .
\label{u}
\end{eqnarray}
{}From the point of view of the holonomy representation, in which the basic
variables are time-independent, $\{u(t),{\bar u}(t)\}$ should simply be
thought of as a useful one-parameter family of commuting operators. The
variables $u$ and $\bar u$ satisfy
\begin{equation}
{d u \over {dt}} =  -  {1\over\alpha} {\csc{2t\over\alpha}}{\bar u},\quad
\hbox{\em or equivalently}\quad
{d u \over {dk}} = - {\alpha\over 4} \sin{2t\over\alpha} {\bar u}.
\label{uut}
\end{equation}
In the $u$ representation, the operators $\hat u$ and ${\hat u}^{\dag}$
will act by multiplication, while suitable linear combinations of the
${\hat r}_1^{\pm}$ will act by differentiation: from (\ref{dc1}),
\begin{eqnarray}
{\hat r}_1^-e^{it/\alpha} + {\hat r}_1^+e^{-{it/\alpha}}
& =& -{2\hbar\over\alpha} (\sin{2t\over\alpha})^{{1\over 2}}
{\partial \over {\partial u}}, \nonumber \\
{\hat r}_1^-e^{-{it/\alpha}} + {\hat r}_1^+e^{it/\alpha}
& =& {2\hbar\over\alpha} (\sin{2t\over\alpha})^{{1\over 2}}
{\partial \over {\partial {\bar u}}} .
\label{uu}
\end{eqnarray}

\section{Relating Representations\label{sec3}}

The ultimate goal of this paper is to relate the quantum theories that
arise from the holonomy and ADM representations, in order to investigate
the role of the modular group in each theory.  To explore this issue,
it is necessary to first understand the classical relationship between
the two approaches.  This requires that we refer back to the space of
classical solutions of (2+1)-dimensional gravity.  For spacetimes with
the topology ${\bf R}\times T^2$ this space is, fortunately, completely
understood.

In the ``proper time gauge'' $N=1$, $N^i=0$, the first-order field
equations
\begin{eqnarray}
R^{ab} &=& d\omega^{ab} - \omega^{ac}\wedge
  \omega_{c}{}^{b} = -\Lambda e^a \wedge e^b \nonumber\\
R^{a} &=& de^a -\omega^{ab}\wedge e_b = 0
\label{bc2}
\end{eqnarray}
are solved by
\begin{eqnarray}
e^0 &=&dt \nonumber\\
e^1&=& {\alpha \over 2} \left[(r_1^+ - r_1^-)dy +(r_2^+ - r_2^-)dx\right]
 \sin{t \over {\alpha}} \\
e^2&=& {\alpha \over 2} \left[(r_1^+ + r_1^-)dy +(r_2^+ + r_2^-)dx\right]
 \cos{t \over {\alpha}} \nonumber
\label{c1}
\end{eqnarray}
\begin{eqnarray}
\omega^{12}&=&0  \nonumber\\
\omega^{01}&=& -{1 \over 2} \left[(r_1^+ - r_1^-)dy +(r_2^+ - r_2^-)dx\right]
 \cos{t \over {\alpha}} \\
\omega^{02}&=& {1 \over 2} \left[(r_1^+ + r_1^-)dy +(r_2^+ + r_2^-)dx\right]
 \sin{t \over {\alpha}} ,\nonumber
\label{c2}
\end{eqnarray}
where $x$ and $y$ each have period $1$.  It is straightforward to check
that the parameters $r_a^\pm$ in (\ref{c1})--(\ref{c2}) are precisely the
parameters (\ref{cc6}) that determine the holonomies.  The York time $k$
for this metric is
\begin{equation}
k = -{d\over dt}\ln\sqrt{{}^{(2)}g} = -{2\over\alpha}\cot{2t\over\alpha} ,
\label{c5}
\end{equation}
which ranges monotonically from $-\infty$ to $\infty$ as $t$ varies from
$0$ to $\pi\alpha/2$, so the slices of constant $t$ are precisely the
slices of constant $k$.

Now, recall that any metric on a constant $k$ slice is diffeomorphic to
one of the form (\ref{bb5}), and that this form defines the ADM variable
$\tau$.  The modulus can thus be read off from the expression (\ref{c1})
for the triad: it is
\begin{equation}
\tau=  \left(r_2^-e^{it/\alpha} + r_2^+e^{-{it/\alpha}}\right)^{\lower2pt%
 \hbox{$\scriptstyle -1$}}\left(r_1^-e^{it/\alpha} +
 r_1^+e^{-{it/\alpha}}\right) .
\label{c3}
\end{equation}
The conjugate variable $p$ can be similarly determined from the canonical
momenta $\pi^{ij}$, which may be computed from (\ref{c1}); one finds that
\begin{equation}
p= {i\alpha\over 2}\csc{2t\over \alpha}\left(r_2^+e^{it/\alpha}
 + r_2^-e^{-{it/\alpha}}\right)^{\lower2pt%
 \hbox{$\scriptstyle 2$}}.
\label{c4}
\end{equation}
{}From (\ref{c3})--(\ref{c4}), the Hamiltonian (\ref{11}) that generates
development in $k$ is
\begin{equation}
H = {\alpha \over {2\sqrt{k^2 - 4\Lambda}}}(r_1^-r_2^+ - r_1^+r_2^-) ,
\label{15}
\end{equation}
while from (\ref{c5}), development in coordinate time $t$ is generated
by
\begin{equation}
H' = {dk \over dt}H= (k^2 - 4\Lambda) H=
 \csc{2t\over\alpha}(r_1^-r_2^+ - r_1^+r_2^-) .
\label {hpr}
\end{equation}

Equations (\ref{c3})--(\ref{hpr}) give us our desired relationship between
the ADM and holonomy representations.  Equivalently, in terms of the
operators $u$ and $\bar u$ defined in the preceding section, we have
\begin{equation}
{\hat\tau} = -{2\hbar\over\alpha} u^{-1}{\partial \over {\partial u}} ,\quad
{\hat\tau}^{\dag} = {2\hbar\over\alpha}{\partial \over {\partial {\bar u}}}
{\bar u}^{-1}
\label{mu}
\end{equation}
and
\begin{equation}
\hat p=  {i\alpha \over 2} {\bar u}^2,\quad
{\hat p}^{\dag}= -{i\alpha \over 2} u^2,
\label{pu}
\end{equation}
whereas the Hamiltonians (\ref{15})--(\ref{hpr}) are
\begin{eqnarray}
\hat H&=&{ {i\alpha\hbar} \over 4}\sin{2t \over {\alpha}}(
\bar u{\partial \over {\partial u}} + u {\partial \over {\partial
\bar u}})\nonumber\\
{\hat H}^{\prime}&=&{ {i\hbar} \over \alpha}\csc{2t \over {\alpha}}(
\bar u{\partial \over {\partial u}} + u {\partial \over {\partial
\bar u}}) .
\label{hh}
\end{eqnarray}
With these orderings, it may be checked that the modulus and momentum
satisfy
\begin{equation}
[{\hat \tau}^\dagger,\hat p]=[{\hat \tau},\hat p^\dagger]= 2i\hbar,
\qquad
  [{\hat \tau},\hat p] = [{\hat \tau}^\dagger,\hat p^\dagger] = 0 ,
\label{dc5}
\end{equation}
in agreement with (\ref{da1}), by virtue of the commutators (\ref{dc1})
of the ${\hat r}_a^\pm$.  Moreover, their time evolution is given by
the standard Heisenberg equations of motion
\begin{equation}
[\hat p, \hat H'] = i\hbar {d\hat p\over dt},
  \qquad  [{\hat \tau}, \hat H'] = i\hbar {d{\hat \tau}\over dt} ,
\label{dc55}
\end{equation}
or equivalently,
\begin{equation}
[ {\hat u}, {\hat H}'] = i \hbar {d {\hat u} \over dt} ,\quad
[ {\hat{\bar u}}, {\hat H}'] = i \hbar {{d {\hat{\bar u}}} \over dt} .
\label{ut}
\end{equation}
In effect, equations (\ref{c3})--(\ref{c4}) can be viewed as the
general four-parameter solution of the quantum mechanical Heisenberg
equations of motion, with the ${\hat r}^\pm_a$ serving as (operator-valued)
parameters.  The action (\ref{28}) of the classical modular group on the
holonomy parameters induces, through (\ref{c3}) and (\ref{c4}), the standard
action (\ref{bb15}) on the torus modulus and momentum, thus confirming
consistency.  The corresponding quantum action is discussed in the next
section.

\section{Modular Transformations in the Holonomy Representation
\label{sec4}}

We have seen that the modular group acts classically on the torus modulus,
momentum, and holonomy parameters as
\begin{equation}
\begin{array}{lllll}
S: & \tau\rightarrow -\tau^{-1} ,\quad
   & p\rightarrow \bar \tau^2 p ,\quad
   & r_1^{\pm}\rightarrow  r_2^{\pm} ,\quad
   & r_2^{\pm}\rightarrow - r_1^{\pm},\\
T: & \tau\rightarrow \tau+1 ,\quad
   & p\rightarrow p ,\quad
   & r_1^{\pm}\rightarrow r_1^{\pm} + r_2^{\pm} ,\quad
   & r_2^{\pm}\rightarrow r_2^{\pm} ,
\end{array}
\label{228}
\end{equation}
and that the transformation (\ref{c3})--(\ref{c4}) between representations
preserves this action.  The goal of this section is to find operators
that generate the quantum version of these transformations.

The simplest starting point is the holonomy representation.  It is easily
checked that the modular transformations of $\hat r_a^\pm$ are generated
by conjugation with the unitary operators\begin{equation}
{\hat T}^{\pm} =
  \exp \left\{\pm {i\alpha \over 2\hbar}({\hat r}_2^{\pm})^2\right\} ,
\label{t}
\end{equation}
\begin{equation}
{\hat S}^{\pm} =
\exp \left\{\pm {i\pi\alpha \over 4\hbar}\left[({\hat r}_1^{\pm})^2
+ ({\hat r}_2^{\pm})^2 \right]\right\} .
\label{s}
\end{equation}
(See the Appendix for a brief description of methods for demonstrating
this and similar relations.)  The first of these appeared in reference
\cite{NR1} in a different notation.  The second was calculated independently
by the two authors, and appeared in \cite{mn1} and \cite{mn2}.  The
operators $\hat T$ and $\hat S$ are related to a set of six constants of
motion  $C_i^{\pm},\ i=1,2,3$, calculated from the holonomies
\cite{mn1}.  These
global constants of motion were first calculated classically, for
$\Lambda=0$, in terms of the ADM modulus and momentum, in \cite{m}.
Explicitly,\footnote{The remaining global constants $C_3^{\pm}= r_1^{\mp}
r_2^{\mp}$ (see \cite{mn1}) are related to $C_1^{\pm}$ and $C_2^{\pm}$ by
$\left\{C_1^{\pm},C_2^{\pm}\right\}=\pm {4 \over \alpha } C_3^{\pm}$.  When
quantized, they generate a scaling $r_1^{\pm}\rightarrow
e^{-\epsilon}r_1^{\pm},\ r_2^{\pm}\rightarrow e^{\epsilon}r_2^{\pm}$.  The
moduli and momenta scale as $\tau \rightarrow e^{-2\epsilon}\tau,\ p
\rightarrow e^{2\epsilon}p$, leaving the commutators (\ref{dc1}) and
(\ref{dc5}) and the Hamiltonian (\ref{15}) invariant.}
\begin{eqnarray}
{\hat T}^{\pm}
&=& \exp \left\{\pm{i\alpha \over 2\hbar}C_2^{\mp}\right\} \nonumber \\
{\hat S}^{\pm}
&=& \exp \left\{\pm{i\pi\alpha \over 4\hbar}(C_1^{\mp} + C_2^{\mp})\right\} .
\label{c}
\end{eqnarray}

We next consider the induced action of $\hat S$ and $\hat T$ on the
modulus and momentum, expressed in the operator ordering given by
equations (\ref{c3}) and (\ref{c4}).  Note first that while the classical
transformations (\ref{bb15}) of $\tau$ translate easily into operator
language, the $S$ transformation of $p$ involves potential ordering
ambiguities.  The ordering (\ref{c3}) that we are considering here
corresponds to a transformation
\begin{equation}
S: \hat p\rightarrow  {{\hat \tau}^{\dag} \over 2}
({\hat \tau}^{\dag} \hat p + \hat p {\hat \tau}^{\dag}) ,
\label{s1}
\end{equation}
while the choice
$\tau=  \left(r_1^-e^{it/\alpha} + r_1^+e^{-{it/\alpha}}
\right)\left(r_2^-e^{it/\alpha} + r_2^+e^{-{it/\alpha}}\right)^{\lower2pt%
 \hbox{$\scriptstyle -1$}}$,
for example, would have led to
\[
S: \hat p\rightarrow
({\hat \tau}^{\dag} \hat p + \hat p {\hat \tau}^{\dag})
{{\hat \tau}^{\dag} \over 2},
\]
the cases differing from each other and the classical limit by terms of
order $\hbar$.  For both orderings the commutators (\ref{dc5}) are invariant
and the identities (\ref{modid}) are satisfied.

Quantum mechanically, we can use equation (\ref{c4}) to reexpress $\hat T$
in terms of $\hat p$ and its adjoint.  We obtain
\begin{equation}
{\hat T}= {\hat T}^+{\hat T}^-
= \exp\left\{ {i\over2\hbar} ({\hat p}+ {\hat p}^\dagger)\right\}
\label{t1}
\end{equation}
Using the commutators (\ref{dc5}), we easily see that conjugation by this
operator generates the transformation (\ref{228}) of $\hat\tau$ and $\hat p$.
The $S$ transformation is more complicated, but the operator (\ref{s})
can also be expressed in terms of the modulus and momentum: using
(\ref{c3}) and (\ref{c4}), we eventually obtain
\begin{eqnarray}
{\hat S}= {\hat S}^+{\hat S}^-
&=& \exp\left\{ {i\pi \over {8\hbar }} \left[
2({\hat p}^{\dag} + \hat p) + {\hat\tau}^{\dag}({\hat\tau}^{\dag}\hat p
+ \hat p {\hat\tau}^{\dag})
+ ({\hat\tau}{\hat p}^{\dag} + {\hat p}^{\dag} {\hat \tau}){\hat \tau}
\right]\right\}\nonumber\\
&=&\exp\left\{ {i\pi \over {4\hbar }} \left[
{\hat p}^{\dag} + \hat p + ({\hat\tau}^{\dag})^2 \hat p
+ {\hat p}^{\dag} ({\hat \tau})^2 + i\hbar({\hat\tau}^{\dag}-
{\hat\tau})
\right]\right\} ,
\label{ss1}
\end{eqnarray}
which differs from the classical expression \cite{mn1} by terms of order
$\hbar$. It can be shown, with some difficulty, that the operator
(\ref{ss1}) generates the desired transformations (\ref{228}) for $\hat\tau$
and (\ref{s1}) for $\hat p$ (see Appendix).

Finally, note that in the $u$ representation, the operators (\ref{t1}) and
(\ref{ss1}) become
\begin{equation}
\hat T=\exp \left\{ {\alpha\over{4\hbar}}\left[{\bar u}^2 - u^2\right]\right\}
\label{tu}
\end{equation}
and
\begin{equation}
\hat S=\exp \left\{-{{\pi\alpha} \over{8\hbar}}\left[
{{4{\hbar}^2}\over {{\alpha}^2}}{{\partial^2} \over {\partial {u}^2}}
-u^2
-{{4{\hbar}^2}\over {{\alpha}^2}}{{\partial^2} \over {\partial {\bar u}^2}}
+ {\bar u}^2\right]\right\}.
\label{su}
\end{equation}

\section{The Transformation between Representations \label{sec5}}

In section \ref{sec2}, we described two quantizations of (2+1)-dimensional
gravity in the torus universe ${\bf R}\times T^2$.  At first sight, the ADM
representation looks like a standard ``Schr{\"o}dinger picture'' quantum
theory, with time-dependent states whose evolution is determined by a
Hamiltonian operator.  The holonomy representation is  more mysterious,
but it resembles a ``Heisenberg picture'' quantum theory, characterized by
time-independent states and time-dependent operators.  This description
suggests that there should be a unitary transformation between the two
representations, which could help in the interpretation of both.

One way to construct such a transformation is to start in the Heisenberg
picture and diagonalize the generalized position operators ${\hat q}_{H}(t)$
for all $t$---that is, to find a family of wave functions $K(x,t)$ such
that for any given $t$, $K(x,t)$ is an eigenfunction of ${\hat q}_{H}(t)$
with eigenvalue $x$.  (The suffixes $H$ and $S$ stand for ``Heisenberg''
and ``Schr{\"o}dinger.'')  Consider, for example, a free particle of mass
$m$ in one spatial dimension.  The Heisenberg states are functions
$\psi_{H}(x_0)$ of an initial position $x_0$, and the position operator
is
\begin{equation}
{\hat q}_{H}(t)
  = {\hat q}_{H}(0) + {t\over m}{\hat p}_{H}(0)
  = x_0 - {{i\hbar t}\over m}{\partial\ \over\partial x_0} .
\label{T1}
\end{equation}
A simple computation shows that the eigenstates
\begin{equation}
{\hat q}_{H}(t) K(x,t|x_0) = x K(x,t|x_0)
\label{T2}
\end{equation}
are
\begin{equation}
K(x,t|x_0) = \left( {m\over 2\hbar\pi t}\right)^{1/2} \exp \left\{
  -{im\over{2\hbar t}}(x-x_0)^2 \right\} .
\label{T3}
\end{equation}
The exponent in (\ref{T3}) is determined by equation (\ref{T2}); the
prefactor is fixed by the normalization requirement that
\begin{equation}
\int dx_0\, K^*(x,t|x_0)K(x',t|x_0) = \delta(x-x') .
\label{T3a}
\end{equation}
It is now easily checked that the complex conjugate kernel $K^*(x,t|x_0)$
satisfies the free particle Schr{\"o}dinger equation,
\begin{equation}
-{{\hbar^2}\over 2m}{{\partial^2 K^*} \over {\partial x^2}} =
i \hbar {{\partial K^*} \over {\partial t}}
\end{equation}
and that a general
Schr{\"o}dinger wave function can be written as a superposition
\begin{equation}
{\tilde\psi}_{S}(x,t)
  = \int dx_0\, K^*(x,t|x_0) \psi_{H}(x_0) .
\label{T4}
\end{equation}
Equation (\ref{T4}) implies that
\begin{equation}
\langle \phi_H|{\hat q}_H(t)|\psi_H\rangle
= \int dx\, \phi_S^*(x,t)x\psi_S(x,t) ,
\label{T4a}
\end{equation}
as  required for a transformation between representations.  In fact, the
kernel $K^*(x,t|x_0)$ is just the standard propagator for a free particle,
and equation (\ref{T4}) is simply the time evolution of the state
$\psi_{H}(x_0)$, considered as an initial state in the Schr{\"o}dinger
picture.

In (2+1)-dimensional quantum gravity, the analogous kernel can be
obtained by diagonalizing the operators $\hat\tau_1$ and $\hat\tau_2$,
or equivalently $\hat\tau$ and $\hat\tau^\dagger$.  In the $u$
representation of equations (\ref{mu})--(\ref{pu}), we thus require
that
\begin{eqnarray}
{\hat\tau}K(\tau,\bar\tau,t|u,{\bar u}) &=&
-{2\hbar \over\alpha}u^{-1}
{\partial\ \over\partial u}K(\tau,\bar\tau,t|u,{\bar u})
  = \tau K(\tau,\bar\tau,t|u,{\bar u}) \nonumber\\
{\hat\tau}^\dagger K(\tau,\bar\tau,t|u,{\bar u}) &=&
{2\hbar \over\alpha}{\partial\ \over\partial {\bar u}}\left[ {\bar u}^{-1}
  K(\tau,\bar\tau,t|u,{\bar u}) \right]
  = {\bar\tau} K(\tau,\bar\tau,t|u,{\bar u}) ,
\label{T5}
\end{eqnarray}
where $\tau$ and $\bar\tau$ are eigenvalues.  It is easily checked that
the solution is
\begin{equation}
K(\tau,\bar\tau,t|u,{\bar u})
  = {\alpha\tau_2\over2\pi\hbar}{\bar u}(t)\exp\left\{
  -{\alpha\over 4\hbar}\tau u(t)^2
  + {\alpha\over 4\hbar}{\bar\tau}{\bar u}(t)^2
  \right\} .
\label{T6}
\end{equation}
The prefactor in equation (\ref{T6}) is again determined by normalization:
we demand that
\begin{equation}
\int du_1du_2\,K^*(\tau,\bar\tau,t|u,{\bar u})K(\tau',\bar\tau',t|u,{\bar u})
  = \tau_2{}^2\delta(\tau_1-\tau_1')\delta(\tau_2-\tau_2') .
\label{T7}
\end{equation}
(The integration measure $du_1du_2$ is equal to $dr_2{}^+dr_2{}^-$, with
no additional Jacobian, so the integral (\ref{T7}) is compatible with our
original choice of variables in section \ref{sec2}.  The delta function
on the right-hand side of (\ref{T7}) is the one appropriate for the
Weil-Petersson metric (\ref{bb13a}) on Teichm{\"u}ller space.)

By analogy with equation (\ref{T4}), our candidates for ``Schr{\"o}dinger
picture'' wave functions are therefore
\begin{eqnarray}
{\tilde\psi}(\tau,\bar\tau,t) &=&
  \int du_1du_2 K^*(\tau,\bar\tau,t|u,{\bar u})\psi(u,{\bar u}) \nonumber\\
\psi(u,{\bar u}) &=&
  \int {d^2\tau\over\tau_2{}^2} K(\tau,\bar\tau,t|u,{\bar u})
  {\tilde\psi}(\tau,\bar\tau,t) .
\label{T8}
\end{eqnarray}
These integrals are not yet well-defined, however: we have not specified
the region of integration, and as we saw in section \ref{sec2}, the
proper choice of ``Heisenberg picture'' wave functions $\psi(u,{\bar u})$
requires a better understanding of the action of the modular group.  In
the next section, we will use equation (\ref{T8}) to define Heisenberg
picture wave functions.  For the moment, let us treat (\ref{T8}) as a
formal expression.

Our ultimate goal is to understand the modular transformations of
$\psi(u,{\bar u})$.  An obvious starting point is to investigate the
actions of the operators $\hat S$ and $\hat T$ of the preceding section
on $K(\tau,\bar\tau,t|u,{\bar u})$.  In the $u$ representation, $\hat T$
acts by multiplication, and it is easy to see from (\ref{tu}) and (\ref{T6})
that
\begin{equation}
{\hat T}K(\tau,\bar\tau,t|u,{\bar u}) = K(\tau+1,\bar\tau+1,t|u,{\bar u}) .
\label{T9}
\end{equation}
The $T$ transformations thus act in the standard way on the modulus $\tau$.
The action of $\hat S$ is rather more complicated to work out, since from
(\ref{su}), $\hat S$ is now a differential operator.  It is safest to work
with real variables $r_2^\pm$ and their conjugates, or for simplicity
with rescaled variables
\begin{eqnarray}
x = \sqrt{{\alpha \over \hbar}}r_2^+ , \qquad
-i{\partial\ \over\partial x} = \sqrt{{\alpha \over \hbar}}r_1^+ \nonumber\\
y = \sqrt{{\alpha \over \hbar}}r_2^- , \qquad
i{\partial\ \over\partial y} = \sqrt{{\alpha \over \hbar}}r_1^- ,
\label{T10}
\end{eqnarray}
in terms of which, from (\ref{su}),
\begin{equation}
{\hat S} = \exp\left\{ {\pi i\over4} \left[
  - {\partial^2\ \over\partial x^2} + x^2
  + {\partial^2\ \over\partial y^2} - y^2 \right] \right\} .
\label{T11}
\end{equation}
The action of this operator can be studied by means of a simple trick.
Note first that
\begin{equation}
e^{ikx} = \sqrt{2\pi}\sum_{n=0}^\infty i^n\psi_n(k)\psi_n(x) ,
\label{T12}
\end{equation}
where the $\psi_n$ are normalized harmonic oscillator wave functions
\cite{Mehler}.  These wave functions are eigenfunctions of the differential
operator in the exponent in (\ref{T11}),
\begin{equation}
\left({\partial^2\ \over\partial x^2} - x^2\right) \psi_n(x) =
  -(2n+1)\psi_n(x) ,
\label{T13}
\end{equation}
and thus
\begin{eqnarray}
{\hat S}e^{ikx} &=&
  e^{\pi i/4}\sqrt{2\pi}\sum_{n=0}^\infty (-1)^n\psi_n(k)\psi_n(x)
\nonumber\\
  &=& e^{\pi i/4}\sqrt{2\pi}\sum_{n=0}^\infty\psi_n(k)\psi_n(-x) =
  e^{\pi i/4}\sqrt{2\pi}\delta(x+k) .
\label{T14}
\end{eqnarray}
Similarly,
\begin{equation}
{\hat S}e^{ik'y} = e^{-\pi i/4}\sqrt{2\pi}\delta(y-k') .
\label{T15}
\end{equation}
Now consider an arbitrary function $F(x,y)$ with a Fourier transform
${\tilde F}(k,k')$:
\begin{eqnarray}
F(x,y) &=& {1\over2\pi} \int dkdk'\, e^{ikx}e^{ik'y}{\tilde F}(k,k')
  \nonumber\\
{\tilde F}(u,v) &=& {1\over2\pi} \int dadb\, e^{-iau}e^{-ibv}F(a,b) .
\label{T16}
\end{eqnarray}
Equations (\ref{T14})--(\ref{T15}) then imply that
\begin{equation}
({\hat S}F)(x,y) = {\tilde F}(-x,y) =
{1\over2\pi} \int dadb\, e^{iax}e^{-iby}F(a,b) .
\label{T17}
\end{equation}
The operator $\hat S$ thus acts by Fourier transformation.  In retrospect
this is perhaps not surprising: by (\ref{28}), $\hat S$ interchanges the
observables $r_1^\pm$ with their conjugates $r_2^\pm$, thus acting as a
transformation from ``position space'' to ``momentum space.''

We can now apply (\ref{T17}) to our kernel $K(\tau,\bar\tau,t|u,{\bar u})$.
A straightforward calculation shows that
\begin{equation}
{\hat S}K(\tau,\bar\tau,t|u,{\bar u})
  = -\left( {\tau\over\bar\tau}\right)^{1/2}
  K(-{1\over\tau},-{1\over\bar\tau},t|u,{\bar u}) .
\label{T18}
\end{equation}
Were it not for the phase on the right-hand side, this would be exactly
what we would expect from the standard action (\ref{bb15}) of $S$ on
moduli space.  The phase makes the transformation ``covariant'' rather
than ``invariant.''  This phase is characteristic of modular forms of
weight $-1/2$, which can be viewed as spinors on moduli space
\cite{Fay,Maass,Rankin,Terras}.  Such modular forms have appeared in
previous work on (2+1)-dimensional gravity with $\Lambda=0$
\cite{dirac,ordering}, although with a different representation
of $\hat S$ and $\hat T$.

A similar computation shows that the complex conjugate kernel
$K^*(\tau,\bar\tau,t|u,{\bar u})$ transforms as
\begin{eqnarray}
{\hat S}K^*(\tau,\bar\tau,t|u,{\bar u})
  &=& -\left( {\bar\tau\over\tau}\right)^{1/2}
  K^*(-{1\over\tau},-{1\over\bar\tau},t|u,{\bar u}) , \nonumber\\
{\hat T}K^*(\tau,\bar\tau,t|u,{\bar u})
  &=& K^*(\tau-1,\bar\tau-1,t|u,{\bar u}) ,
\label{kstar}
\end{eqnarray}
characteristic of a modular form of weight $1/2$.  The covariant
Laplacian $\Delta_{1/2}$ for modular forms of weight $1/2$ is the Maass
Laplacian (\ref{da5}) \cite{Fay,Maass,Rankin,Terras},
\begin{equation}
\Delta_{1/2} = - \tau_2{}^2 \left( {\partial^2\ \over\partial\tau_1{}^2}
  + {\partial^2\ \over\partial\tau_2{}^2}\right)
  + i\tau_2 {\partial\ \over\partial\tau_1} + {3\over4} .
\label{T19}
\end{equation}
As noted in section \ref{sec2}, this operator differs from the ordinary
Laplacian $\Delta_0$ by terms of order $\hbar$, and can thus be viewed as
a different operator ordering of the standard Laplacian.  A straightforward
computation now shows that
\begin{equation}
\left( {i\alpha\over2}\sin{2t\over\alpha}\,{\partial\ \over\partial t}
\right)^2 K^*(\tau,\bar\tau,t|u,{\bar u})
= (\Delta_{1/2}-1) K^*(\tau,\bar\tau,t|u,{\bar u}) .
\label{T20}
\end{equation}
Up to a constant order $\hbar$ correction, this is the square of the
reduced phase space Schr{\"o}dinger equation (\ref{da3}) with an operator
ordering appropriate for a form of weight $1/2$, and it serves as a check
that our kernel $K^*(\tau,\bar\tau,t|u,{\bar u})$ behaves as it ought to.
In particular, (\ref{T20}) implies that the ``Schr{\"o}dinger picture''
wave functions of equation (\ref{T8}) will satisfy a similar
Klein-Gordon-like equation.

It is also interesting to consider the action of the ``Heisenberg
picture'' Hamiltonian (\ref{hpr}) on $K^*(\tau,\bar\tau,t|u,{\bar u})$.
{}From (\ref{hh}), we see that
\begin{equation}
{\hat H}^{\prime}K^*(\tau,\bar\tau,t|u,{\bar u})
= -i\hbar {\partial\ \over\partial t} K^*(\tau,\bar\tau,t|u,{\bar u}) .
\label{T24}
\end{equation}
Equations (\ref{T20}) and (\ref{T24}) imply that, in some sense,
${\hat H}^{\prime} \sim (\Delta_{1/2})^{1/2}$, i.e., that the first-order
holonomy-based quantum theory is a ``square root'' of the second-order
ADM theory.  A similar phenomenon was noted earlier in the
theory with $\Lambda=0$, although with different variables \cite{dirac}.
Whether this relation can be made more rigorous remains an open question.
The basic problem is that the square root of a Laplacian is highly
nonunique: it can be defined mode by mode in a spectral decomposition,
but the sign of the square root can be chosen arbitrarily for each mode.
It is not clear which, if any, of this infinite number of square roots
should be associated with ${\hat H}^{\prime}$.

\section{Modular Transformations of Holonomy Wave Functions \label{sec6}}

We are now ready to use the results of the preceding section to
analyze the behavior of the ``Heisenberg picture'' wave function
$\psi(u,{\bar u})$ under modular transformations.  Our strategy will be
to use the well-understood properties of the ``Schr{\"o}dinger picture''
wave function ${\tilde\psi}(\tau,{\bar\tau},t)$, along with the
transformation (\ref{T8}) between representations.

We begin with the second equation in (\ref{T8}),
\begin{equation}
\psi(u,{\bar u}) =
  \int_{{\cal F}} {d^2\tau\over\tau_2{}^2} K(\tau,\bar\tau,t|u,{\bar u})
  {\tilde\psi}(\tau,\bar\tau,t) .
\label{M1}
\end{equation}
Since $K(\tau,\bar\tau,t|u,{\bar u})$ is, roughly speaking, a modular
form of weight $-1/2$, as implied by equation (\ref{T18}), we might expect
${\tilde\psi}(\tau,\bar\tau,t)$ to be a form of weight $1/2$, that is,
a function invariant under the transformations
\begin{equation}
{\hat S}{\tilde \psi}(\tau,{\bar\tau},t) =
-\left({{\bar\tau}\over\tau}\right)^{1/2}
{\tilde \psi}(-{1\over\tau},-{1\over{\bar\tau}},t) , \quad
{\hat T}{\tilde \psi}(\tau,{\bar\tau},t) =
{\tilde \psi}(\tau+1,{\bar\tau}+1,t) .
\label{M1a}
\end{equation}
We shall see below that this is indeed the case.

Our first task is to determine the range of integration $\cal F$ in
(\ref{M1}).  This can be fixed by the requirement that $\psi(u,{\bar u})$
be properly normalized:
\begin{eqnarray}
\int d^2u |\psi(u,{\bar u})|^2 &=&
\int d^2u \int_{{\cal F}}{d^2\tau\over\tau_2{}^2}
\int_{{\cal F}}{d^2\tau'\over\tau'_2{}^2}
K(\tau,{\bar\tau},t|u,{\bar u})K^*(\tau',{\bar\tau}',t|u,{\bar u})
{\tilde\psi}(\tau,{\bar\tau},t){\tilde\psi}^*(\tau',{\bar\tau}',t)
\nonumber\\
&=& \int_{{\cal F}}{d^2\tau\over\tau_2{}^2}
\int_{{\cal F}}{d^2\tau'\over\tau'_2{}^2}
\tau'_2{}^2 \delta^2(\tau-\tau')
{\tilde\psi}(\tau,{\bar\tau},t){\tilde\psi}^*(\tau',{\bar\tau}',t)
= \int_{{\cal F}}{d^2\tau\over\tau_2{}^2}
|{\tilde\psi}(\tau,{\bar\tau})|^2 ,\nonumber\\
\label{M2}
\end{eqnarray}
where we have used the orthonormality relation (\ref{T7}).  But we
understand the normalization of ``Schr{\"o}dinger picture'' wave
functions ${\tilde\psi}(\tau,{\bar\tau},t)$: the right-hand side
of (\ref{M2}) will be unity when $\cal F$ is a fundamental region for
the action (\ref{bb15}) of the modular group on Teichm{\"u}ller space.

With this choice of integration region, we take (\ref{M1}) as the
{\em definition\/} of $\psi(u,{\bar u})$.  Let us now consider the
action of the operators $\hat S$ and $\hat T$ of section \ref{sec4} on
this wave function.  From (\ref{kstar}), it is easy to see that
\begin{eqnarray}
{\hat T}\psi(u,{\bar u}) &=& \int_{{\cal F}}{d^2\tau\over\tau_2{}^2}
({\hat T}K)(\tau,\bar\tau,t|u,{\bar u}){\tilde\psi}(\tau,\bar\tau,t)
\nonumber\\
&=& \int_{{\cal F}}{d^2\tau\over\tau_2{}^2}
K(\tau+1,\bar\tau+1,t|u,{\bar u}){\tilde\psi}(\tau,\bar\tau,t)
\nonumber\\
&=& \int_{{\cal F}}{d^2\tau\over\tau_2{}^2}
K(\tau+1,\bar\tau+1,t|u,{\bar u}){\tilde\psi}(\tau+1,\bar\tau+1,t) ,
\label{M3}
\end{eqnarray}
where the invariance of ${\tilde\psi}$ under the transformations
(\ref{M1a}) has been used in the last line.  Changing integration variables
to $\tau+1$ and ${\bar\tau}+1$, we see that
\begin{equation}
{\hat T}\psi(u,{\bar u}) = \int_{T^{-1}{\cal F}}
 {d^2\tau\over\tau_2{}^2} K(\tau,\bar\tau,t|u,{\bar u})
 {\tilde\psi}(\tau,\bar\tau,t) ,
\label{M4}
\end{equation}
where $T^{-1}{\cal F}$ is the new fundamental region obtained from $\cal F$
by a $T^{-1}$ transformation.  A similar argument shows that
\begin{equation}
{\hat S}\psi(u,{\bar u}) = \int_{S^{-1}{\cal F}}
 {d^2\tau\over\tau_2{}^2} K(\tau,\bar\tau,t|u,{\bar u})
 {\tilde\psi}(\tau,\bar\tau,t) ,
\label{M5}
\end{equation}
provided ${\tilde\psi}(\tau,\bar\tau,t)$ is a modular form of weight
$1/2$, invariant under the transformations (\ref{M1a}).  (The extra phase
factor in (\ref{M1a}) is needed to cancel the phase in the transformation
(\ref{T18}), as anticipated.)

Now, the kernel $K(\tau,\bar\tau,t|u,{\bar u})$ is {\em not\/}
modular invariant, and the shift of integration region in equations
(\ref{M4}) and (\ref{M5}) matters: the wave function $\psi(u,{\bar u})$
is not invariant under the action of the mapping class group.\footnote{This
fact was first pointed out to one of the authors (S.C.) by Jorma Louko.}
Indeed, there is a sense in which $\psi(u,{\bar u})$ and (for example)
${\hat T}\psi(u,{\bar u})$ differ maximally---they are, in fact, orthogonal.
To see this, we can repeat the calculation of equation (\ref{M2}); from the
orthonormality of $K(\tau,\bar\tau,t|u,{\bar u})$, we now obtain
\begin{equation}
\left\langle \psi|{\hat T}\psi\right\rangle =
\int_{T^{-1}{\cal F}}{d^2\tau\over\tau_2{}^2}
\int_{{\cal F}}{d^2\tau'\over\tau'_2{}^2}
\tau'_2{}^2 \delta^2(\tau-\tau')
{\tilde\psi}(\tau,{\bar\tau},t){\tilde\psi}^*(\tau',{\bar\tau}',t) .
\label{M6}
\end{equation}
But the regions $\cal F$ and $T^{-1}{\cal F}$ are disjoint except on a set
of measure zero, so the delta function in (\ref{M6}) is identically zero.

A similar argument shows that if $g$ is any nontrivial modular
transformation, then
\begin{equation}
\left\langle \psi|{\hat g}\psi\right\rangle = 0 .
\label{M7}
\end{equation}
In fact, this conclusion can be strengthened.  Let $\psi_1(u,{\bar u})$
and $\psi_2(u,{\bar u})$ be two different wave functions defined by
integrals of the form (\ref{M1}) over the same fundamental region
$\cal F$.  Repeating the computation of equation (\ref{M6}), we now
see that
\begin{equation}
\left\langle \psi_1|{\hat g}\psi_2\right\rangle = 0
\label{M8}
\end{equation}
for any nontrivial modular transformation $g$.

In accord with the results of references \cite{Louko,Giulini,Peldan},
our ``Heisenberg picture'' wave functions are not modular invariant.
But the ``maximal noninvariance'' of equation (\ref{M8}) is almost as good.
Pick a fundamental region $\cal F$, and consider the set of wave functions
defined by (\ref{M1}).  These will form a subspace ${\cal H}_{{\cal F}}$
of the Hilbert space $\cal H$ of square-integrable functions of $(u_1,u_2)$
or $(r_2^+,r_2^-)$.  A modular transformation $g$ maps this subspace into
an orthogonal subspace ${\cal H}_{g^{-1}{\cal F}}$, which is obtained by
integrals of the form (\ref{M1}) over the translated fundamental region
$g^{-1}{\cal F}$.  In fact, the modular group splits the space $\cal H$
into an infinite set of orthogonal subspaces.

These subspaces are physically equivalent.  Indeed, let $\hat{\cal O}$
be an arbitrary modular invariant operator on $\cal H$.  There is no
reason to expect the orthogonal subspaces to be superselected---that is,
if $\psi\in{\cal H}_{\cal F}$, it need not be the case that ${\hat{\cal O}}
\psi\in{\cal H}_{\cal F}$---so if we wish to restrict ourselves to a single
subspace, we must appropriately project $\hat{\cal O}$ to that subspace.
Let ${\hat P}_{\cal F}$ denote the standard Hilbert space projector onto
the subspace ${\cal H}_{{\cal F}}$, and define $\hat{\cal O}_{\cal F} =
{\hat P}_{\cal F} \hat{\cal O}{\hat P}_{\cal F}$.  Then if $\psi_1,\psi_2
\in{\cal H}_{\cal F}$, it follows that
\begin{eqnarray}
\langle\psi_1|{\hat{\cal O}}_{{\cal F}}|\psi_2\rangle &=&
\langle\psi_1|{\hat P}_{\cal F}{\hat{\cal O}}{\hat P}_{\cal F}|\psi_2\rangle
\nonumber\\ &=&
\langle\psi_1|{\hat g}^{-1}{\hat P}_{g{\cal F}}{\hat g}{\hat{\cal O}}
{\hat g}^{-1}{\hat P}_{g{\cal F}}{\hat g}|\psi_2\rangle
\nonumber\\ &=&
\langle g\psi_1|{\hat P}_{g{\cal F}}{\hat{\cal O}}{\hat P}_{g{\cal F}}
|g\psi_2\rangle =
\langle g\psi_1|{\hat{\cal O}}_{g{\cal F}}|g\psi_2\rangle ,
\label{M9}
\end{eqnarray}
where we have used the modular invariance of $\hat{\cal O}$ and the
fact that ${\hat g}^{-1}{\hat P}_{g{\cal F}}{\hat g} = {\hat P}_{\cal F}$.
Matrix elements can thus be computed in any of the subspaces
${\cal H}_{g{\cal F}}$, and the appropriate restrictions of modular
invariant operators will give the same physics.

The modulus $\hat\tau$, of course, is not an invariant operator, and
its matrix elements will depend on the choice of subspace.  But this
is not surprising, since the same is true classically.  One can build
invariant operators from $\hat\tau$, whose matrix elements satisfy
(\ref{M9}).  One example is the operator version of the modular function
$J(\tau)$ of Dedekind and Klein \cite{Rankin},
\begin{equation}
J(\tau) = {\left(60 G_4(\tau)\right)^3 \over
  \left(60 G_4(\tau)\right)^3 - 27\left(140 G_6(\tau)\right)^2} ,
\label{e15}
\end{equation}
where the $G_{2k}(\tau)$ are Eisenstein series,
\begin{equation}
G_{2k}(\tau) = {\sum_{m,n\in\mathbf{Z}}}^\prime {1\over (m+n\tau)^{2k}} .
\label{e16}
\end{equation}
(The prime means that the value $m=n=0$ is excluded from the sum.)
It may be shown that any meromorphic modular function is a rational
function of $J(\tau)$.  Such functions are certainly less familiar than
trigonometric functions, but in principle they are no more extraordinary.
Since $J(\tau)$ depends only on the modulus and not the momentum, there
are no ordering ambiguities, and (\ref{e15}) may be taken to be an
operator expression.

What we have discovered is a ``quantum mechanical fundamental region''
for the modular group.  As several authors have pointed out
\cite{Louko,Giulini,Peldan}, the modular group does not act
nicely (that is, properly discontinuously) on the configuration space
of the first-order formalism.  But we now see that the modular group
{\em does\/} act nicely on the corresponding Hilbert space, which is all
that is required for a sensible quantum theory.

\section{Conclusion}

The phase space of (2+1)-dimensional gravity with $\Lambda<0$ on a
manifold ${\bf R}\times\Sigma$ has two natural descriptions: as the cotangent
bundle of the Teichm{\"u}ller space of $\Sigma$, and as a space of
$\hbox{SL}(2,{\bf R})\otimes\hbox{SL}(2,{\bf R})$ holonomies.  Classically, the
two descriptions are equivalent,\footnote{Strictly speaking, one must
restrict the holonomies to ensure that the metric is nonsingular and
that the slices $\Sigma$ are spacelike \cite{Louko}.} and can be viewed
as different coordinate choices for a single space.  On the {\em phase
space}, the mapping class group has a properly discontinuous action
in either set of coordinates, and invariant functions are well defined.

To quantize such a phase space, however, one must choose a polarization,
that is, a distinction between ``positions'' and ``momenta.''  Therein
lies the root of the problem discussed in references
\cite{Louko,Giulini,Peldan}.
While the mapping class group acts nicely on the phase space, there is
no guarantee that it does so on the configuration space, and hence no
assurance that one can find invariant wave functions.  When $\Sigma$ is
a torus, this is precisely what goes wrong: the modular group fails to
act properly discontinuously on a ``configuration space'' of holonomies,
and the definition of invariant wave functions becomes highly problematic.

One could, of course, evade this issue by choosing a different polarization
\cite{Peldan2}.  But the polarization for which the problems arise is a
natural one, and it seems implausible that a perfectly good choice of
classical coordinates should lead to such disastrous consequences for
the quantum theory.

In this paper, we have solved this problem.  By constructing the exact
transformation between the ADM and holonomy states, we have shown that
the modular group {\em does\/} have a nice, albeit unexpected, action
on the holonomy states.  There are, indeed, no invariant wave functions
in the holonomy representation.  Instead, the modular group acts
on the Hilbert space in much the same way that it acts on Teichm{\"u}ller
space---it splits the Hilbert space into physically equivalent orthonormal
``fundamental regions,'' each one of which is equivalent to the Hilbert
space that arises from ADM quantization.  In the course of our argument,
we have also derived a collection of explicit operator representations of
the torus mapping class group.

The splitting of the Hilbert space described in section \ref{sec6} relies
on the transformation (\ref{T8}) between representations, and thus refers
back to the ADM quantum theory.  It would be desirable to have a description
that depended only on the intrinsic properties of the Hilbert space in the
holonomy representation.  We do not yet have such a description, but we
see no reason why one should not exist.

\vspace{1.5ex}
\begin{flushleft}
\large\bf Acknowledgements
\end{flushleft}
This work was supported in part by National Science Foundation grant
PHY-93-57203, Department of Energy grant DE-FG03-91ER40674, the European
Commission programmes HCM CHRX-CT93-0362 and TMR ERBFMRX-CT96-0045,
and INFN Iniziativa Specifica TO 10 (FI 41).  We would also like to
thank an anonymous referee, whose comments helped clarify several issues.

\appendix
\section{}

The generators (\ref{t})--(\ref{s}) and (\ref{t1})--(\ref{ss1}) of
modular transformations act by conjugation, and to compute their
action, one must evaluate expressions of the form
\begin{equation}
e^{A} B e^{-A}= B + [A,B] + {[A,[A,B]] \over {2!}} + \dots
\label{bch}
\end{equation}
In this appendix, we briefly describe two ways to evaluate such
expressions, one based on explicit summation and a second based on
a trick that converts the problem to one of solving differential
equations.

As an example of the explicit calculation, write the generator $\hat S$
of equation (\ref{ss1}) as
\begin{equation}
\hat S = \exp\left\{ {\pi i\over8} (\hat a+{\hat a}^{\dag}) \right\} ,
\label{aa1}
\end{equation}
where
\begin{eqnarray}
\hbar{\hat a} &=& 2{\hat p}^\dagger + {\hat\tau}{\hat p}^\dagger{\hat\tau}
 + {\hat p}^\dagger{\hat\tau}^2,\nonumber\\
\hbar{\hat a}^{\dag} &=& 2{\hat p}
 + {\hat\tau}^{\dag}{\hat p}{\hat\tau}^{\dag}
 + ({\hat\tau}^{\dag})^2{\hat p}
\label{a0}
\end{eqnarray}
and
\begin{equation}
[\hat a, {\hat a}^{\dag}] = 0 .
\end{equation}
To use (\ref{bch}) to evaluate the transformation of ${\hat p}^{\dag}$,
for example, one must compute the multiple commutators
\begin{equation}
[A,[A,[A,......[A, {\hat p}^{\dag}]....]]]
\label{a5}
\end{equation}
where $A=-{\pi i\over 8}{\hat a}$.  Note first that by (\ref{dc5}),
\begin{equation}
[\hat a,\hat\tau] = 4i(1+{\hat\tau}^2) , \quad
[\hat a,{\hat p}^\dagger] = -8i{\hat p}^\dagger{\hat\tau} - 4\hbar .
\label{a2}
\end{equation}
Direct computation shows that the odd commutators are all proportional:
\begin{equation}
[A,[A,[A,......[A, {\hat p}^{\dag}]....]]]_{2n+1}=
(-\pi^2)^n[A,{\hat p}^{\dag}] .
\label{a6}
\end{equation}
Similarly, the even commutators can be computed to be
\begin{equation}
[A,[A,[A,......[A, {\hat p}^{\dag}]....]]]_{2n}
= (-\pi^2)^{n-1}[A,[A,{\hat p}^{\dag}]]
= { {(-1)^{n-1}{\pi^{2n}}}\over 4}(\hbar \hat a-4{\hat p}^{\dag}) .
\label{a7}
\end{equation}
It follows that the sum of the odd commutators in (\ref{bch}) is
\begin{equation}
[A,{\hat p}^{\dag}]\sum_{n=0}^\infty{{(-{\pi}^2)^n} \over {(2n+1)!}}=
{1\over \pi}[A,{\hat p}^{\dag}] \sum_{n=0}^\infty{{(-1)^n {\pi}^{2n+1} }
\over {(2n+1)!}}=
{1 \over \pi}[A,{\hat p}^{\dag}]\sin{\pi}=0 ,
\label{a8}
\end{equation}
whereas the sum of the even commutators, starting from
$[A,[A,{\hat p}^{\dag}]]$, is
\begin{equation}
\left({\hbar{\hat a}\over 4}-{\hat p}^{\dag}\right)
\sum_{n=1}^\infty {{(-1)^{n-1}{\pi^{2n}}} \over {(2n)!}}
=\left({\hbar{\hat a}\over 4}-{\hat p}^{\dag}\right)(1-\cos{\pi})
= {\hbar{\hat a}\over 2}-2{\hat p}^{\dag} .
\label{a9}
\end{equation}
Thus from (\ref{bch}),
\begin{equation}
e^{A} {\hat p}^{\dag} e^{-A}= {\hat p}^{\dag}
+ \left({\hbar{\hat a}\over 2}-2{\hat p}^{\dag}\right)
= {1\over2}({\hat p}^\dagger{\hat\tau}^2
+ {\hat\tau}{\hat p}^\dagger{\hat\tau}) ,
\end{equation}
in agreement with (\ref{s1}), as required.

We next give an alternative method for calculating the $S$ transformations
of $\hat\tau$ and $\hat p$.  Let
\begin{equation}
F(s) = e^{-is\hat a}{\hat\tau} e^{is\hat a} , \quad
G(s) = e^{-is\hat a}{\hat p}^\dagger e^{is\hat a} .
\end{equation}
By (\ref{aa1}), the transformed values of $\hat\tau$ and ${\hat p}^\dagger$
are simply $F(\pi/8)$ and $G(\pi/8)$.  But by differentiating  $F(s)$
and $G(s)$ with respect to $s$ and using the commutators (\ref{a2}), we
can reduce the problem to one of solving a pair of differential equations,
\begin{equation}
{dF\over ds} = 4(1+F^2) , \quad
{dG\over ds} = -8GF + 4i\hbar .
\label{a3}
\end{equation}
The first equation in (\ref{a3}) has the solution
\begin{equation}
F(s) = \tan 4(s-s_0) ,
\end{equation}
with initial conditions
\begin{equation}
F(0) = -\tan 4s_0 = \hat\tau .
\end{equation}
Hence
\begin{equation}
F(\pi/8) = \cot 4s_0 = -{\hat\tau}^{-1} ,
\label{a4}
\end{equation}
yielding the correct transformation (\ref{228}) for $\hat\tau$. To
calculate the corresponding transformation of ${\hat p}^{\dag}$, observe
that by (\ref{a3}),
\begin{equation}
{d\ \over ds}[G(1+F^2)] = -8GF(1+F^2) + 4i\hbar
(1+F^2) + 2GF{dF\over ds}  = 4i\hbar(1+F^2) = i\hbar{dF\over ds} ,
\end{equation}
and thus
\begin{equation}
G(s)(1+F(s)^2) - G(0)(1+F(0)^2) = i\hbar(F(s)-F(0)) .
\end{equation}
Setting $s=\pi/8$ and using (\ref{a4}), we find that
\begin{equation}
G(\pi/8) = {\hat p}^\dagger{\hat\tau}^2 - i\hbar{\hat\tau} =
{1\over2}({\hat p}^\dagger{\hat\tau}^2
+ {\hat\tau}{\hat p}^\dagger{\hat\tau}) ,
\end{equation}
recovering (\ref{s1}).


\begin{thebibliography}{99}
\bibitem{Carlip_bk} S.~Carlip, {\it Quantum Gravity in 2+1 Dimensions\/}
 (Cambridge University Press, Cambridge, 1998).
\bibitem{HosNak} A.~Hosoya and K.~Nakao, Class.\ Quant.\ Grav.\ {\bf 7},
 163 (1990); Prog.\ Theor.\ Phys. {\bf 84}, 739 (1990).
\bibitem{Mon} V.~Moncrief, J.\ Math.\ Phys.\ {\bf 30}, 2907 (1989).
\bibitem{Fujiwara} Y.~Fujiwara and J.~Soda, Prog.\ Theor.\ Phys. {\bf 83},
 733 (1990).
\bibitem{Achu} A.~Ach{\'u}carro and P.~K.~Townsend, Phys.\ Lett.\ {\bf B180},
 89 (1986).
\bibitem{Witten} E.~Witten, Nucl.\ Phys.\ {\bf B311}, 46 (1988/89).
\bibitem{observables} S.~Carlip, Phys.\ Rev.\ {\bf D42}, 2647 (1990).
\bibitem{dirac} S.~Carlip, Phys.\ Rev.\ {\bf D45}, 3584 (1992).
\bibitem{ordering} S.~Carlip, Phys.\ Rev.\ {\bf D47}, 4520 (1993).
\bibitem{NR2} J.~E.~Nelson and T.~Regge, Nucl.\ Phys.\ {\bf B328}, 190
 (1989).
\bibitem{NR1} J.~E.~Nelson and T.~Regge, Phys.\ Lett.\ {\bf B272},
 213 (1991).
\bibitem{NR3} J.~E.~Nelson and T.~Regge, Commun.\ Math.\ Phys.\ {\bf 141},
 211 (1991).
\bibitem{NR5} J.~E.~Nelson and T.~Regge, in {\em Integrable Systems and
 Quantum Groups, Pavia 1990}, edited by M.~ Carfora,  M.~Martellini, and
 A.~Marzuoli (World Scientific, Singapore, 1992).
\bibitem{NR4} J.~E.~Nelson and T.~Regge, Commun.\ Math.\ Phys.\ {\bf 155},
 561 (1993).
\bibitem{NR0} J.~E.~Nelson and T.~Regge, Phys.\ Rev.\ {\bf D50}, 5125
 (1994).
\bibitem{NRZ} J.~E.~Nelson, T.~Regge and F.~Zertuche, Nucl.\ Phys.\
 {\bf B339}, 516 (1990).
\bibitem{Unruh} W.~G.\ Unruh and P.\ Newbury, Int.\ J.\ Mod.\ Phys.\
 {\bf D3}, 131 (1994).
\bibitem{CarlipNelson} S.~Carlip and J.~E.~Nelson, Phys.\ Rev.\
{\bf D51}, 5643 (1995).
\bibitem{CarlipNelson2} S.~Carlip and J.~E.~Nelson, Phys.\ Lett.\
{\bf B324}, 299 (1994).
\bibitem{Puzio} R.~Puzio, Class.\ Quant.\ Grav.\ {\bf 11}, 609 (1994).
\bibitem{Louko} J.~Louko and D.~M.\ Marolf, Class.\ Quant.\ Grav.\ {\bf 11},
 311 (1994).
\bibitem{Giulini} D.~Giulini and J.~Louko, Class.\ Quant.\ Grav.\ {\bf 12},
 2735 (1995).
\bibitem{Peldan} P.~Peld{\'a}n, Phys.\ Rev.\ {\bf D53}, 3147 (1996).
\bibitem{York} J.~W.~York, Phys.\ Rev.\ Lett.\ {\bf 28}, 1082 (1972).
\bibitem{Fay} J.~D.~Fay, J.~Reine\ Angew.\ Math.\ {\bf 293}, 143  (1977).
\bibitem{Maass} H.\ Maass, {\it Lectures on Modular Functions of One Complex
 Variable\/} (Tata Institute, Bombay, 1964).
\bibitem{Rankin} R.~A.~Rankin, {\it Modular Forms and Functions\/}
 (Cambridge University Press, Cambridge, 1977).
\bibitem{Terras} A.~Terras, {\it Harmonic Analysis on Symmetric Spaces
 and Applications I\/} (Springer, New York, 1985).
\bibitem{Fischer} See, for example, A.~E.\ Fischer and V.~Moncrief, in
 {\em Global Structure and Evolution in General Relativity}, edited
 by S.~Cotsakis and G.~W.\ Gibbons (Springer, New York, 1996); Gen.\ Rel.\
 Grav.\ {\bf 28}, 221 (1996).
\bibitem{mn1} V.~Moncrief and J.~E.~Nelson, Int.~J.\ Mod.\ Phys.\
 {\bf D6}, 5 (1997).
\bibitem{mn2} V.~Moncrief and J.~E.~Nelson, in {\em Proceedings of Eighth
 Marcel Grossmann Meeting on General Relativity, Jerusalem 1997}, to be
 published by World Scientific.
\bibitem{m} V.~Moncrief, J.~Math.\ Phys.\ {\bf 31}, 2978 (1990).
\bibitem{Mehler} F.~G.\ Mehler, J.~Reine Angew.\ Math.\ {\bf 66}, 161
 (1866); see also {\em The Bateman Manuscript Project}, edited by A.~Erdelyi
 (McGraw-Hill, New York, 1954), Vol.~3, Chap.~19.
\bibitem{Peldan2} P.~Peld{\'a}n, Class.\ Quant.\ Grav.\ {\bf 13}, 221
 (1996).

\end{thebibliography}
\end{document}